%% file: iclr2024_conference.tex
\documentclass{article} 
\usepackage{GEM_workshop_2024,times}

\input{math_commands.tex}

\usepackage[utf8]{inputenc} 
\usepackage[T1]{fontenc}    
\usepackage{hyperref}       
\usepackage{url}            
\usepackage{booktabs}       
\usepackage{amsfonts}       
\usepackage{nicefrac}       
\usepackage{microtype}      
\usepackage{graphicx}
\usepackage{amsmath}
\usepackage{tabularx}
\usepackage{wrapfig}
\usepackage{multirow}
\usepackage{threeparttable}
\usepackage{hyperref}
\usepackage{bm}
\usepackage{enumitem}
\usepackage{subfigure}
\usepackage{algpseudocode}
\newcommand{\angstrom}{\textup{\AA}}
\usepackage[ruled,linesnumbered,vlined]{algorithm2e}

\newcommand{\model}{F$^3$low}

\title{F$^3$low: Frame-to-Frame Coarse-grained Molecular Dynamics with SE(3) Guided Flow Matching}

\author{%
  Shaoning Li$^{1,2}$$\,$\thanks{Equal contribution.},\,\, Yusong Wang$^{3,4}$$\,$\footnotemark[1]$\,$,\,\, Mingyu Li$^{5}$$\,$\footnotemark[1],\,\, Jian Zhang$^{5}$, \, Bin Shao$^{3}$,\\ \textbf{Nanning Zheng$^{4}$,\,\, Jian Tang$^{1,6,7}$\thanks{Corresponding author.} }\\
  $^1$ Mila - Québec AI Institute 
  $^2$ Université de Montréal 
  $^3$ Microsoft Research AI4Science \\ 
  $^4$ National Key Laboratory of Human-Machine Hybrid Augmented Intelligence, \\ 
  National Engineering Research Center for Visual Information and Applications, \\
  and Institute of Artificial Intelligence and Robotics, Xi'an Jiaotong University \\ 
  $^5$ Medicinal Chemistry and Bioinformatics Center, Shanghai Jiao Tong University School of Medicine \\
  $^6$ HEC Montréal 
  $^7$ Canadian Institute for Advanced Research (CIFAR)
}

\iclrfinalcopy 
\begin{document}

\maketitle

\begin{abstract}
Molecular dynamics (MD) is a crucial technique for simulating biological systems, enabling the exploration of their dynamic nature and fostering an understanding of their functions and properties.
To address exploration inefficiency, emerging enhanced sampling approaches like coarse-graining (CG) and generative models have been employed.
In this work, we propose a \underline{Frame-to-Frame} generative model with guided \underline{Flow}-matching (\textbf{\model}) for enhanced sampling, which
(a) extends the domain of CG modeling to the $\operatorname{SE}(3)$ Riemannian manifold; 
(b) retreating CGMD simulations as autoregressively sampling guided by the former frame via flow-matching models;
(c) targets the protein backbone, offering improved insights into secondary structure formation and intricate folding pathways.
Compared to previous methods, \model~allows for broader exploration of conformational space.
The ability to rapidly generate diverse conformations via force-free generative paradigm on $\operatorname{SE}(3)$ paves the way toward efficient enhanced sampling methods.
\end{abstract}

\section{Introduction}



\begin{figure}[htbp]
    \centering
    \includegraphics[width=\textwidth]{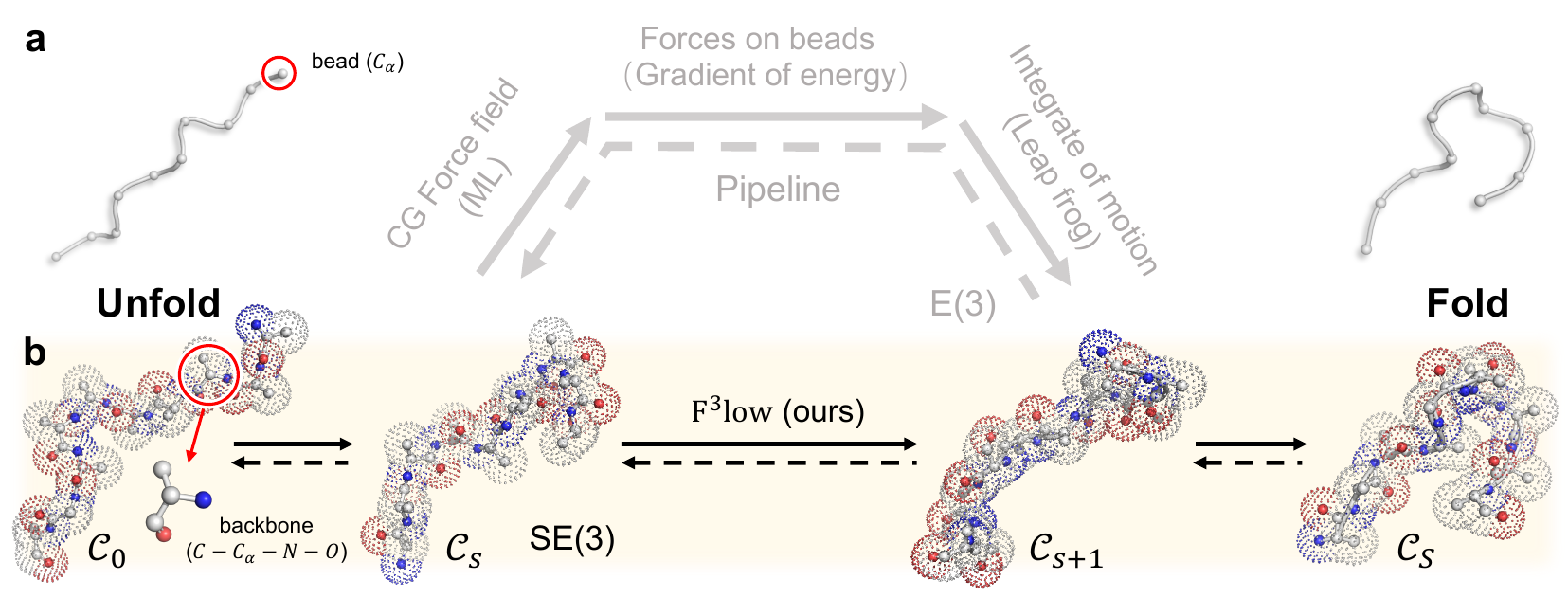}
    \caption{
        Overview of the enhanced sampling methods, traditional CGMD pipeline (panel \textbf{a}) and generative models (e.g., \model, panel \textbf{b}).
        Traditional CGMD relies on empirical forces applied to each CG bead to calculate the next frame.
        In contrast, generative models operate by directly sampling the next frame from a prior distribution, bypassing the need for explicit force calculations.
    }
    \label{fig:cgflow_paradigm}
\end{figure}

Molecular dynamics (MD) is a computational technique that simulates the motion and behavior of biological macromolecules systems without the need for wet lab experiments.
By employing physics-based force fields and algorithms, MD provides valuable insights into the dynamic aspects of protein structures, such as protein folding event (\cite{lindorff2011fast}, Fig.~\ref{fig:cgflow_paradigm}).
This approach is instrumental in elucidating the functions and interactions of proteins at atomic level.

However, MD simulations faces challenges in efficiently exploring the conformation space due to the difficulty of crossing the high energy-barriers, which requires milliseconds of simulation steps and renders infeasible. 
To address this issue, \textit{enhanced sampling} methods are developed to encourage the feasible exploration of conformation space via MD simulations.
One of the enhanced sampling methods is \textit{coarse-grained} molecular dynamics (CGMD) (\cite{kmiecik2016coarse}).
It involves creating a coarse-grained representation by slicing out atoms from each residue up to its $C_{\alpha}$, termed as \textit{beads} shown in Fig.~\ref{fig:cgflow_paradigm}(a). 
In this process, atomistic forces generated by potential functions or machine learning force fields are applied to beads on the CG space to perform simulations. 
This facilitates the smoothing of energy landscapes, preventing the occurrence of traps in local minima.
Another emerging method utilizes \textit{probabilistic generative models} to generate conformations targeting the Boltzmann distribution, without the dependency of empirical forces and motion integration (Fig.~\ref{fig:cgflow_paradigm}(b)).
\cite{klein2023timewarp} employs normalizing flow to map structures of small peptides from Gaussian noises for the next step, highlighting transferability between states with relatively larger timesteps.

In this paper, we leverage the advantages of both CGMD and generative models, presenting \model, for enhanced sampling.
Firstly, we extend the common modeling resolution of CGMD from $C_{\alpha}$ level to \textit{backbone} level ($C-C_{\alpha}-N-O$, Fig.~\ref{fig:cgflow_paradigm}(b)).
In essence, the space of conformation poses is expanded from $\mathbb{R}^{3N}$ (or $\operatorname{E}(3)^{N}$) to $\operatorname{SE}(3)^{N}$, with $N$ denoting the number of residues.
In such way, it preserves the structural geometry, enabling the direct observation of the secondary structure and folding pathway without the need for any reconstruction.
Secondly, we leverage the guided flow on the SE(3) manifold (\cite{lipman2022flow, chen2023riemannian, zheng2023guided}) to perform the simulation in a backbone-resolution.
Concretely, as illustrated in Fig.~\ref{fig:cgflow_paradigm}(b), \model~samples the next frame $\mathcal{C}_{s+1}$ from prior distribution (e.g., zero center of mass Gaussian) while being guided by the preceding frame $\mathcal{C}_{s}$ at step $s$.
We term this sampling based on conditional probability as \textit{frame-to-frame} simulation.
This process can be repeated for $S$ steps to conduct complete simulations.
Compared with \cite{klein2023timewarp}, \model~is not confined to small peptides (under 4AA) but can handle larger fast folding proteins (\cite{lindorff2011fast}) spanning from Chignolin (10AA) to Homeodomain (54AA) benefiting from the advantage of coarse-graining.

\section{Methods}

Given a conformation frame $\mathcal{C}_s$ at step $s$, we leverage a probabilistic generative model $P_{\theta}$ to sample the next frame $\mathcal{C}_{s+1}$ guided by $\mathcal{C}_s$ from a prior source distribution $z$:
\begin{equation}
\label{eq:setup}
    \mathcal{C}_{s+1} \sim P_{\theta}(\mathcal{C}_{s+1}; z, \mathcal{C}_{s}).
\end{equation}
This iterative process, starting from an initial frame $\mathcal{C}_0$ and repeated $S$ times, results in the generation of a complete trajectory $\Gamma = [\mathcal{C}_0, \dots, \mathcal{C}_S]$. 
We refer to this procedure based on conditional probability as \textit{frame-to-frame} simulation illustrated in Fig.~\ref{fig:cgflow_paradigm}(b).
After the simulations, we can conduct various trajectory analysis of protein dynamics, including comparing free energy surfaces and calculating transition probabilities between meta-stable states.

\subsection{SE(3) guided flow matching}
In this context, we provide a brief introduction to the general concept of Riemannian flow matching, followed by its extension to the SE(3) space. 
Then we introduce the basic concept of guided flow.

\textbf{From Riemannian to SE(3) flow matching.}
We consider complete connected, smooth \textit{Riemannian manifolds} $\mathcal{M}$ endowed with \textit{metric} $g$.
At each point $x \in \mathcal{M}$ can attach to a \textit{Tangent space} $\mathcal{T}_x$ and the metric $g$ is defined as inner product over $\mathcal{T}_x\mathcal{M}$.
The \textit{continuous normalizing flow} (CNF) $\psi_t : \mathcal{M} \rightarrow \mathcal{M}$ is defined as the solution of the following ordinary differential equation (ODE) along with a time-dependent vector field $u_t \in \mathcal{T}_x\mathcal{M}$: $\frac{d}{dt}\psi_{t}(x) = u_t(\psi_{t}(x))$, with initial conditions $\psi_{0}(x) = x$.
Density distribution on Riemannian manifolds is continuous non-negative functions $p: \mathcal{M} \rightarrow \mathbb{R}_{+}$ the integrate to $\int_{\mathcal{M}}p(x)dx=1$.
Given two distributions $p_0$ and $p_1$ on $\mathcal{M}$, we can define a \textit{probability path} $p_{t}: t \in [0, 1]$ as their interpolation.
In other word, $p_t$ is a sequence of probability distributions generated by vector field $u_t$ through pushing forward $p_0$ to $p_1$.
To learn a CNF,
\textit{flow-matching} (FM) is proposed (\cite{albergo2022building, lipman2022flow}) as a simulation-free method by regressing vector field $u_t$ with a parametric one $v_{\theta}$.
Since computing $u_t$ is intractable and cannot be directly targeted, we always adopt \textit{conditional flow matching} (CFM), attaching with a \textit{conditional probability path} $p_{t}(x|x_1)$ satisfying $p_{0}(x|x_1)=p_{0}(x)$ and $p_1(x|x_1) \approx \delta_{x_1}(x)$.
$\delta_{x_1}(x)$ is the delta probability concentrating all its mass at $x_1$.
Subsequently, the unconditional $p_t$ can be recovered by calculating marginal distribution $p_t = \int_{\mathcal{M}}p_{t}(x|x_1)p_{1}(x_1)dx_1$.
The objective of training CFM is to minimize $v_{\theta}^t$ with conditional vector field $u_t(x|x_1)$.
When plugging the conditional flow $x_t = \psi_t(x_0|x_1)$ with initial condition $\psi_0(x_0|x_1) = x_0$ into the CFM loss, we can reparameterize $u_t(x|x_1)$ with $\frac{d}{dt}\psi_t(x)$,
and rewrite the final CFM loss as:
\begin{equation}
    \mathcal{L}_{\operatorname{CFM}}(\theta) = \mathbb{E}_{t, p_1(x_1), p_0(x_0)} \left \| v_{\theta}^t(x_t) - \dot{x_t} \right \|_g^2,
\end{equation}
where $t \sim \mathcal{U}(0, 1)$, $\dot{x_t} = \frac{d}{dt}x_t = u_t(x|x_1)$ and $\left \| \cdot \right \|_g^2$ is the norm induced by Riemannian metric $g$.

The basic structure of protein backbone is characterized by a rigid transformation $T = (r, x) \in \operatorname{SE}(3)$ (\cite{jumper2021highly}), where this transformation can be decomposed into a rotation $r \in \operatorname{SO}(3)$ and translation vector $x \in \mathbb{R}^3$.
Following \cite{yim2023se, yim2023fast}, the metric $g$ on $\operatorname{SO}(3)$ is $\langle a, a^{\prime} \rangle_{\operatorname{SO}(3)} = \operatorname{Tr}(a a^{\prime\top}) / 2$ where $a$ is the vector in tangent space of $\operatorname{SO}(3)$, referring to $\mathfrak{so}(3)$, and $\operatorname{Tr}(\cdot)$ denotes the trace of the matrix.
The metric $g$ on $\mathbb{R}^3$ is $\langle y, y^{\prime} \rangle_{\mathbb{R}^3} = \sum_{i=1}^3 y_iy_i^{\prime}$ where $y \in \mathbb{R}^3$.
Combining these two metrics, we derive the metric on $\operatorname{SE}(3)$ as $\langle (a,y), (a^{\prime}, y^{\prime}) \rangle_{\operatorname{SE}(3)} = \langle a, a^{\prime} \rangle_{\operatorname{SO}(3)} + \langle y, y^{\prime} \rangle_{\mathbb{R}^3}$.
It allows us construct the conditional flow $\psi_t$ connecting source distribution $p_0$ and target distribution $p_1$.
Given $T_0(r_0, x_0) \sim p_0$ and $T_1(r_1, x_1) \sim p_1$, $\psi_t(T_0|T_1)$ is denoted as the geodesic interpolant path represented on $\operatorname{SO}(3)$ and $\mathbb{R}^3$ respectively:
\begin{equation}
\label{eq:geodesic}
    \begin{aligned}
        r_t &= \exp_{r_0}(t\log_{r_0}(r_1)) \\
        x_t &= (1 - t)x_0 + tx_1, \\
    \end{aligned}
\end{equation}
where $\log$ is logarithm map converting elements in $\operatorname{SO}(3)$ to $\mathfrak{so}(3)$, and $\exp$ is exponential map for inverting the logarithm map.
Referring to \cite{yim2023fast}, the objective can be written as directly predicting the target:
\begin{equation}
    \mathcal{L}_{\operatorname{CFM}-\operatorname{SE}(3)}(\theta) = \mathbb{E}_{t, p_0(T_0), p_1(T_1)}\frac{1}{(1-t)^2} \left[ \left \| \hat{x}_1 - x_1 \right \|_{\mathbb{R}^3}^2 + \left\| \log_{r_t}(\hat{r}_1) - \log_{r_t}(r_1) \right\|_{\operatorname{SO}(3)}^2 \right].
\end{equation}
where $\hat{r}_1$ and $\hat{x}_1$ are the prediction of $v_{\theta}^t(T_t)$.

\textbf{Guided flow.}
Guided flow is first introduced by \cite{zheng2023guided}, which is an adaptation of classifier-free guidance \cite{ho2022classifier} to FM models for conditional generation.
Given the condition $y$, we can extend the unconditional $p_t$ to $p_t(x|y) = \int_{\mathcal{M}}p_t(x|x_1)p_1(x_1|y)dx_1$ and corresponding vector field $u_t(x|y) = \int_{\mathcal{M}} u_t(x|x_1)\frac{p_t(x|x_1)p_1(x_1|y)}{p_t(x)}dx_1$.
Following \cite{zheng2023guided}, the guided vector field can be derived from the combination of $p_1(x)^{1-\omega}p_1(x|y)^{\omega}$:
\begin{equation}
    \tilde{u}_t(x|y) = (1 - \omega)u_t(x) + \omega u_t(x|y).
\end{equation}
where $\omega \in \mathbb{R}$ is a manually selected weight.
\cite{zheng2023guided} also showcases that $u_t(x|y)$ is related to score function $\nabla \log p_t(x|y)$ by $u_t(x|y) = a_t x + b_t \nabla \log p_t(x|y)$, with $a_t$ and $b_t$ as schedulers \footnote{\cite{zheng2023guided} only showcases the guided flow on $\mathbb{R}^{3}$. We will add the proof of its generalization on $\operatorname{SO}(3)$ in the future work.}.

\subsection{\model~for coarse-grained molecular dynamics}
In this part, we introduce our methods, \model, which provides a new perspective on coarse-grained molecular dynamics by iteratively generating conformation frames in a trajectory using a guided flow-matching model.
We outline the basic concept, elaborate on integrating conformation guidance into the flow-matching models based on initial guesses, and provide an overview of the training and sampling procedures.

\textbf{Basic concept.}
As defined in Eq.~\ref{eq:setup}, we leverage a trained guided flow-matching model to sample the next frame.
To be consistent with the aforementioned definitions, we denote $p_0 := z$, $p_1 := \mathcal{C}_{s+1}$ and $y := \mathcal{C}_{s}$.
The conformation $\mathcal{C}$ is modeled with coarse-graining at the protein backbone level, consisting of 4 heavy atoms $C-C_{\alpha}-N-O$.
Let $N$ denote the number of residue within $\mathcal{C}$, the space of conformation poses is $\operatorname{SE}(3)^N$,  where each residue is mapped by a rigid transformation $T_i$ with $i \in [N]$.
And hence $\mathcal{C} = [T_0, \dots, T_N] \in \operatorname{SE}(3)^N$.
To preserve the $\operatorname{SE}(3)$ symmetries, all the translations are carried out within the zero center of mass (CoM) subspace.
This ensures that the prior source distribution $z^x$ and the intermediate $x_t$ remain $\operatorname{SE}(3)$-invariant, and the vector field $u_t$ maintains $\operatorname{SE}(3)$-equivariance.
Here we denote the prior distribution in Eq.~\ref{eq:setup} as $z = [z^r, z^x]$ where $z^x$ represents the Gaussian distribution on $\mathbb{R}^3$ subtracted the CoM. Following \cite{yim2023fast, watson2023novo}, $z^r$ serves as the isotropic Gaussian distribution $\mathcal{IG}_{\operatorname{SO}(3)}$.

It is noteworthy that we \textit{cannot} directly define a flow from $\mathcal{C}_s$ to $\mathcal{C}_{s+1}$ since they belong to the same trajectory and, therefore, are under the same Boltzmann distribution, rather than two separate distributions.
Instead, we formulate the problem as a guided flow, where we sample $\mathcal{C}_{s+1}$ guided by $\mathcal{C}_s$ from a normal prior distribution. 
This aligns with the definition of flow in this context.

\textbf{Integration of conformation guidance.}
As demonstrated in Eq.~\ref{eq:geodesic}, $r_t$ and $x_t$ represent the linear interpolation of $z$ and $\mathcal{C}_{s+1}$.
Then these values are utilized as the input for $v_{\theta}^t$, which predicts the target $\hat{r}_1$ and $\hat{x}_1$.
In the context of classifier-free guidance,  incorporating the condition $y$ (the former conformation $\mathcal{C}_{s}$) into the FM model poses challenges in the case of protein structures.
Unlike image or text labels, representing 3D protein structures is non-trivial. 
Various attempts have been proposed, including methods such as template embedding (\cite{jumper2021highly}) or treating the structures as \textit{initial guesses} before recycling (\cite{bennett2023improving}).
Motivated by \cite{bennett2023improving}, we also consider $\mathcal{C}_{s}$ as a form of initial guess, and integrate it into $z$ as a weighted average in space:
\begin{equation}
    \begin{aligned}
        \tilde{r}_0 &= \exp_{r_0}(\gamma\log_{r_0}(z^r)) \\
        \tilde{x}_0 &= (1 - \gamma) z^x + \gamma x_0,
    \end{aligned}
\end{equation}
where $\gamma$ is a hyperparameter termed as initial guess weight.
During training, it is necessary to compute the optimal transport (OT) between $x_1$ and $z^x$.
And hence the derivation of $\tilde{x}_0$ can be re-written:
\begin{equation}
    \tilde{x}_0 = (1 - \gamma) \operatorname{OT}(z^x, x_1) + \gamma x_0.
\end{equation}
Then we subtract the CoM of $\tilde{x}_0$ to ensure equivariance.
Such linear interpolation helps better preserve the structure information while keeping the model architecture unchanged.
Detailed training and sampling procedures are provided in Appendix Algorithm \ref{algm:training} and Algorithm \ref{algm:sampling}.

\section{Experiment}

\subsection{Free Energy Surface}
To validate the capabilities of the F$^3$low model for conformational space exploration, we conducted training and sampling procedures (see details on Appendix~\ref{appendix:details}) on three representative fast-folding proteins: Chignolin, a $\beta$-hairpin structure; Trpcage, dominated by its dominant $\alpha$-helix; and Homeodomain, a complex assembly of three helices. Further comparative analysis with the coarse-grained machine learning force field (CG-MLFF) simulations~(\cite{majewski2023machine}), which focus solely on the $C_{\alpha}$ atom, has demonstrated the superior performance of our approach in accurately depicting secondary structures by considering all backbone atoms.

\begin{wraptable}{r}{0.6\textwidth}
\centering
\caption{Minimum RMSD value with respect to the crystal structure for all simulations.}  
\label{table:min_rmsd}
\begin{threeparttable}
\resizebox{0.55\textwidth}{!}{
\begin{tabular}{lcccc}  
\toprule
& & Reference & CG-MLFF & CG-F$^3$low \\    
\midrule 
\multirow{2}{*}{Chignolin} & $C_\alpha$ & 0.15 & 0.17 & 0.14 \\  
& Backbone & 0.27 & 0.89 & 0.36 \\  
\midrule 
\multirow{2}{*}{Trpcage} & $C_\alpha$ & 0.45 & 1.03 & 0.56 \\  
& Backbone & 0.58 & 2.41 & 0.62 \\  
\midrule  
\multirow{2}{*}{Homeodomain} & $C_\alpha$ & 0.56 & 2.17 & 0.51 \\  
& Backbone & 0.54 & 2.11 & 0.53 \\  
\bottomrule
\end{tabular}
}
\end{threeparttable}
\end{wraptable}
   
For the purpose of dimensionality reduction, we applied time-lagged independent component analysis~(\cite{perez2013identification}, TICA) to both the CG-MLFF and CG-F$^3$low simulation trajectories. This analysis projected the data onto the leading three or four components, utilizing covariance matrices derived from the reference MD simulations. Additionally, we constructed Markov state models~(\cite{husic2018markov}, MSMs) and applied a reweighting technique to the free energy surfaces, leveraging the stationary distribution of the MSMs. Comprehensive details regarding the TICA and MSM parameters for each protein are provided in the Appendix~\ref{appendix: tica_and_msm}.  
   
\begin{figure}[htbp]
    \centering
    \includegraphics[width=\textwidth]{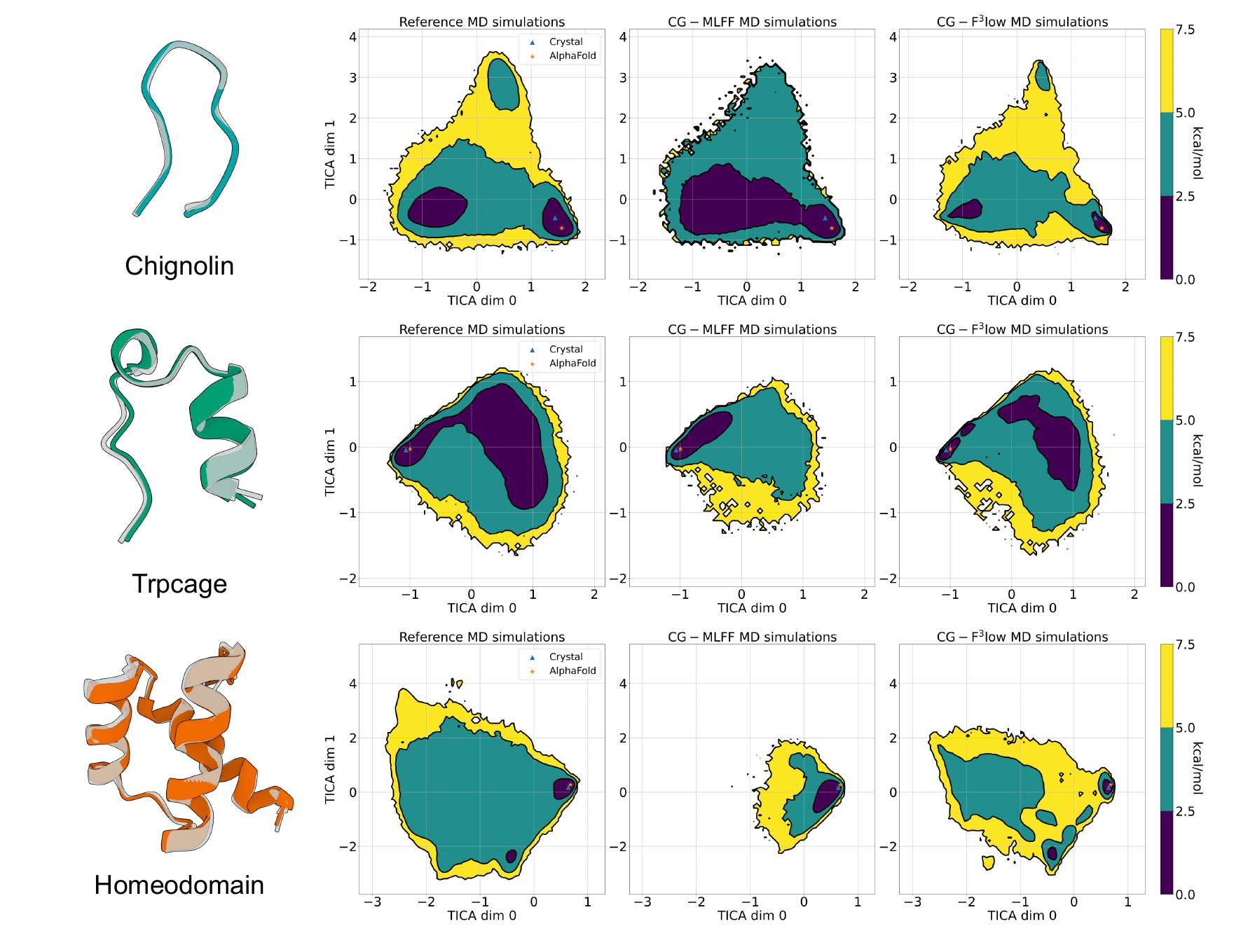}
    \caption{Comparison of free energy surface across reference, CG-MLFF and CG-F$^3$low simulations. 
    The crystal structure (gray) and the lowest RMSD structure in CG-F$^3$low simuluation (colored) are presented on the left, respectively.
    The corresponding Min. RMSD can be found in Table \ref{table:min_rmsd}.}
    \label{fig:energy_surface}
\end{figure}

As illustrated in Fig. \ref{fig:energy_surface}, our evaluation of two relatively simple proteins, Chignolin and Trpcage, demonstrates that both models achieve commendable coverage of the free energy surface. Notably, F$^3$low presented a more precise energy distribution in comparison to the reference data. In the case of the larger protein, Homeodomain, we observed that the CG-MLFF simulations rapidly converged to the native structure and did not sufficiently sample the unstructured conformations. In contrast, F$^3$low effectively explored these less-defined regions, achieving this with a reduced number of starting points (8 as opposed to 32). A broader exploration of the conformational space allows us to go even further in our in-depth analysis of Homeodomain, see Appendix~\ref{appendix:homeodomain_analysis} for more details.
   
Furthermore, we have superimposed the crystal structure and the AlphaFold-predicted~(\cite{jumper2021highly}) structure onto the free energy surface plots. The AlphaFold predictions align remarkably well with the actual crystal structure. Nevertheless, it represents a singular structure. The generative models, on the other hand, are capable of exploring a vastly expanded conformational space, thereby offering more profound insights into the dynamic behavior of proteins.

\begin{figure}[htbp]
    \centering
    \includegraphics[width=\textwidth]{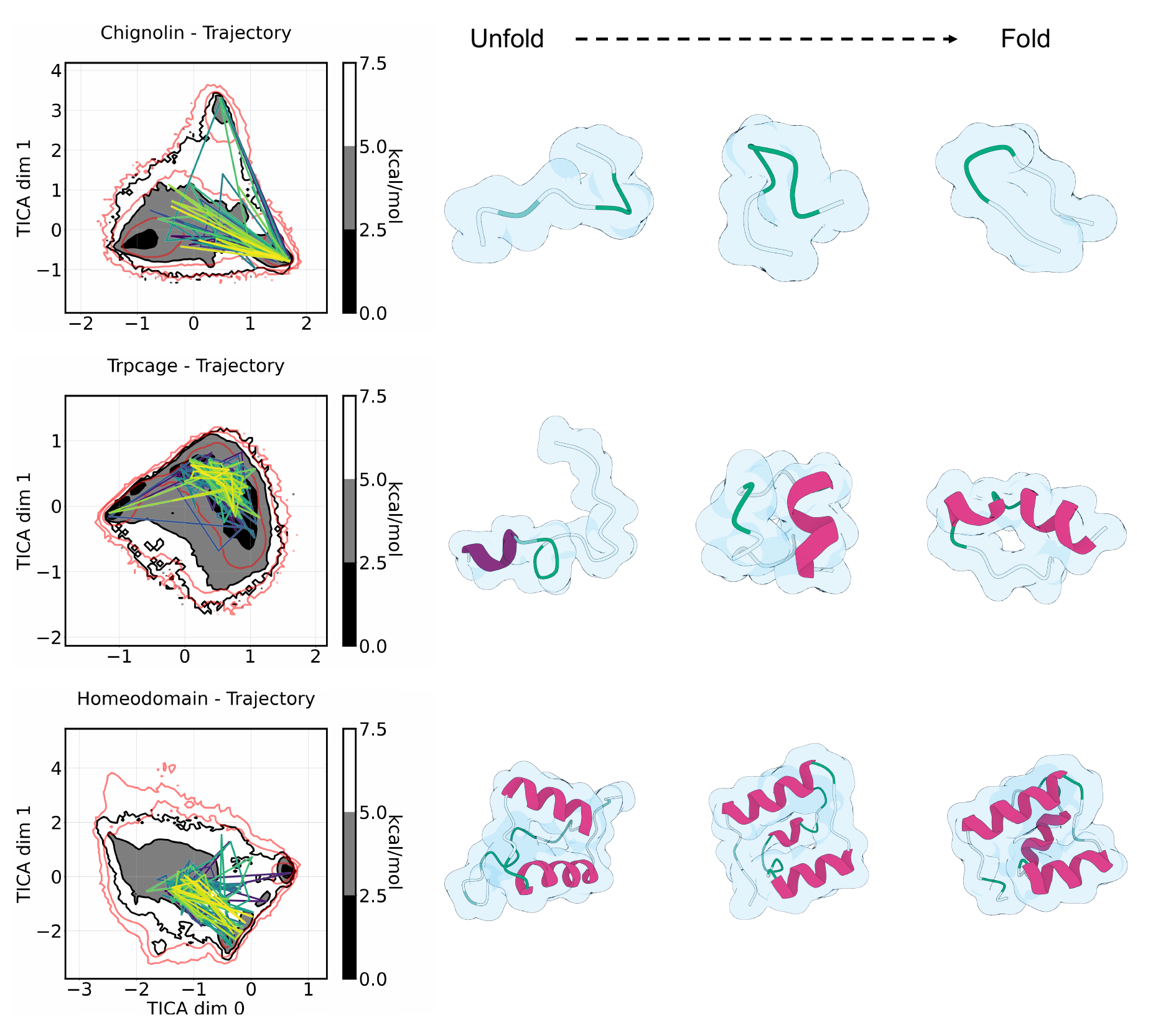}
    \caption{Individual trajectory visualization for Chignolin, Trpcage and Homeodomain. In each visual representation, the simulation trajectory traverses the free energy landscape, depicted in a gradient from purple to yellow.}
    \label{fig:traj_analysis_2}
\end{figure}

We also visualize an individual simulation trajectory for each protein in Fig.~\ref{fig:traj_analysis_2}.
The frames from diluted trajectories imply that our approach is capable of exploring a broader space of potential energy surfaces within a single trajectory, as opposed to solely depending on trajectory ensemble.

\subsection{Comparison of simulated and crystal structures.}

All simulations successfully identified the location of the global minimum on the free energy surface across all proteins. 
Subsequently, we conducted a comparative analysis of the minimum root-mean-square deviation (RMSD) values relative to the crystal structures, as detailed in Table~\ref{table:min_rmsd}.
Our CG-F$^3$low simulations consistently sampled structures with lower RMSD values than CG-MLFF simulations and showed comparable RMSD to the reference simulations.
The efficacy of our algorithm is particularly pronounced when comparing in terms of the backbone. In contrast, CG-MLFF simulations require additional steps to reconstruct the backbone, highlighting the inherent advantages of our approach.

\section{Conclusion}
In this work, we present a novel enhanced sampling method named \model, which is a frame-to-frame generative model with guided flow matching on $\operatorname{SE}(3)$.
\model~extends the modeling domain to $\operatorname{SE}(3)$ with a backbone level-resolution, providing improved insights into secondary structure information and intricate folding pathways.
By analyzing the simulation trajectories of 3 representative proteins and comparing with traditional MLFF, we illustrate the \model's capacity for broadly exploring conformational spaces within a generative paradigm.
In our future work, we intend to incorporate the physical analysis of the simulation trajectories, and conduct the remaining fast folding proteins.
Additionally, the flow process on side-chain torsion angles would be included to perform all-atom protein molecule dynamics.

\newpage
\section*{Acknowledgement}
We thank the reviewers of GEM workshop for their valuable comments.
We thank Adrià Pérez, the author of \cite{majewski2023machine} for providing the detailed settings of free energy surface.
We thank Le Zhuo for valuable discussions of flow matching methods.

This paper is supported by National Key Research and Development Program of China (2023YFF1205103), National Natural Science Foundation of China (81925034) and National Natural Science Foundation of China (62088102).
This paper is also supported by the Natural Sciences and Engineering Research Council (NSERC) Discovery Grant, the Canada CIFAR AI Chair Program, collaboration grants between Microsoft Research and Mila, Samsung Electronics Co., Ltd., Amazon Faculty Research Award, Tencent AI Lab Rhino-Bird Gift Fund, a NRC Collaborative R\&D Project (AI4D-CORE-06) as well as the IVADO Fundamental Research Project grant PRF-2019-3583139727.

\bibliography{iclr2024_conference}
\bibliographystyle{iclr2024_conference}

\newpage
\appendix

\section{Training and Sampling}
We describe the precise training and simulation process in Algorithm~\ref{algm:training} and \ref{algm:sampling} over distribution in $\operatorname{SE}(3)^{N}$.
Different from \cite{yim2023se}, \textit{frame} in this paper is the snapshot within a trajectory, rather than the spatial structure of the protein backbone.
In the training process line 7, when combining the initial guess, an alignment for $x_0$ to $x_1$ is performed before adding it to the transported $z^x$. This step is implemented to maintain training stability.
In the simulation process, the notation $\operatorname{ODEStep}$ (line 10) refers to the Euler method.

\begin{algorithm}
\caption{\model~training on $\operatorname{SE}(3)$}
\label{algm:training}
\SetKwInOut{Input}{Input}\SetKwInOut{Output}{Output}
\Input{Trajectory dataset $\mathcal{C}_{\Gamma}$, flow network $v_{\theta}$, initial guess weight $\gamma$, probability of unconditional training $p_{\operatorname{uncond}}$}
\Output{Trained \model~$v_{\theta}$}
\While{Training}{
    $t \sim \mathcal{U}(0, 1)$\;
    $\mathcal{C}_{s}, \mathcal{C}_{s+1} \sim \mathcal{C}_{\Gamma}$\;
    $z^r, z^x \sim \mathcal{IG}_{\operatorname{SO}(3)}, \; \mathcal{N}(0, 1)$\;
    $z^x \leftarrow z^x - \operatorname{CoM}(z^x)$\;
    \If{Uniform$(0, 1.0) > p_{\operatorname{uncond}}$}{
        $\tilde{r}_0, \tilde{x}_0 \leftarrow \exp_{r_0}(\gamma\log_{r_0}(z^r)), \; (1 - \gamma) \operatorname{OT}(z^x, x_1) + \gamma x_0$\;
        $\tilde{x}_0 \leftarrow \tilde{x}_0 - \operatorname{CoM}(\tilde{x}_0)$\;
    }
    \Else{
        $\tilde{x}_0 \leftarrow \operatorname{OT}(z^x, x_1)$\;
    }
    $r_t, x_t \leftarrow \exp_{r_0}(t\log_{r_0}(r_1)), \; (1 - t)\tilde{x}_0 + tx_1$\;
    $\hat{r}_1, \hat{x}_1 \leftarrow v_{\theta}(r_t, x_t | r_0, x_0)$\;
    $\mathcal{L}_{\operatorname{CFM}-\operatorname{SE}(3)} \leftarrow \frac{1}{(1-t)^2} \left[ \left \| \hat{x}_1 - x_1 \right \|_{\mathbb{R}^3}^2 + \left\| \log_{r_t}(\hat{r}_1) - \log_{r_t}(r_1) \right\|_{\operatorname{SO}(3)}^2 \right]$\;
    $\theta \leftarrow \operatorname{Update}(\theta, \nabla_{\theta}\mathcal{L}_{\operatorname{CFM}-\operatorname{SE}(3)})$\;
}
\Return{$v_{\theta}$}
\end{algorithm}

\begin{algorithm}
\caption{Coarse-grained frame-to-frame simulation on $\operatorname{SE}(3)$ via \model}
\label{algm:sampling}
\SetKwInOut{Input}{Input}\SetKwInOut{Output}{Output}
\Input{Trained \model~$v_{\theta}$, initial start frame $\mathcal{C}_{0}$, guidance parameter $\omega$, initial guess weight $\gamma$, number of ODE steps $n_{\operatorname{ode}}$, number of simulation steps $S$}
\Output{Trajectory $\mathcal{C}_{\Gamma}^S$}
$\mathcal{C}_{\operatorname{cur}} \leftarrow \mathcal{C}_{0}$ \;
$\mathcal{C}_{\Gamma}^S \leftarrow List[\mathcal{C}_0]$\;
$h \leftarrow \frac{1}{n_{\operatorname{ode}}}$\;
\For{$s$ in $[1, \dots, S]$}{
$r_0, x_0 \leftarrow \mathcal{C}_{\operatorname{cur}}$\;
$\tilde{x}_0 \leftarrow (1 - \gamma) z^x + \gamma x_0$\;
$\tilde{r}_0 \leftarrow \exp_{r_0}(\gamma\log_{r_0}(z^r))$\;
    \For{$t=0, h, \dots, 1-h$}{
        $\tilde{v}_{\theta}(r_t, x_t) \leftarrow (1 - \omega) v_{\theta}(r_t, x_t) + \omega v_{\theta}(r_t, x_t | \tilde{r}_0, \tilde{x}_0)$\;
        $x_{t+h} \leftarrow$ ODEStep $(\tilde{v}_{\theta}(r_t, x_t), x_t)$\;
    }
$\mathcal{C}_{s} \leftarrow r_1, x_1$\;
Append$(\mathcal{C}_{\Gamma}^S, \mathcal{C}_{s})$\;
}
\Return{$\mathcal{C}_{\Gamma}^S$}
\end{algorithm}

\newpage
\section{Markov state model estimation}
\label{appendix: tica_and_msm}

Markov state models (MSMs) were constructed for the reference MD simulations, as well as for the CG-MLFF and CG-F$^3$low simulations. For the reference simulations, in line with the methodologies outlined in previous research~\cite{majewski2023machine}, simulation data were transformed into pairwise $C_\alpha$ distances (excluding the terminal five residues of Trpcage). TICA was then employed to reduce the dimensionality of the featurized data, capturing the first 4, 3, and 3 principal components for Chignolin, Trpcage, and Homeodomain, respectively. 
The resulting components were clustered using the K-means algorithm, and this discretized data was utilized for MSM estimation. The CG-MLFF and CG-F$^3$low simulations underwent an identical process. However, for these simulations, covariance matrices derived from the reference MD simulations were used during the TICA projection to ensure a consistent basis for comparison of the first 3 or 4 components. This step was crucial for assessing how accurately the coarse-grained simulations reproduce the free energy surfaces of proteins.  
To enhance the interpretability of the MSMs, the PCCA algorithm was applied to consolidate the microstates into macrostates. The comparative free energy surfaces were generated by partitioning the first two TICA components into an $80 \times 80$ grid and adjusting them through reweighting using the MSM stationary distribution. The detailed parameters are listed in the table below.

\begin{table*}[htbp]  
\centering  
\caption{Parameters of the TICA and MSM analysis for reference, CG-MLFF and CG-F$^3$low simulations.}  
\begin{threeparttable}
\resizebox{\linewidth}{!}{
\begin{tabular}{lcccccc}  
\toprule
& & TICA lag time (steps) & \# of TICA projection components & \# of K-means clusters & MSM lag time (ns) & \# of MSM macrostates \\    
\midrule 
\multirow{3}{*}{Reference} & Chignolin & 20 & 4 & 1200 & 20 & 3 \\  
& Trpcage & 100 & 3 & 500 & 10 & 2 \\   
& Homeodomain & 100 & 3 & 600 & 10 & 3 \\   
\midrule 
\multirow{3}{*}{CG-MLFF} & Chignolin & / & 3 & 200 & 0.001 & 3 \\  
& Trpcage & / & 3 & 200 & 0.01 & 2 \\
& Homeodomain & / & 3 & 200 & 0.01 & 3 \\
\midrule  
\multirow{3}{*}{CG-F$^3$low} & Chignolin & / & 4 & 1200 & 20 & 3 \\  
& Trpcage & / & 3 & 500 & 10 & 2 \\    
& Homeodomain & / & 3 & 600 & 10 & 3 \\     
\bottomrule
\end{tabular}
}
\end{threeparttable}
\end{table*} 



\newpage
\section{Structural analysis of Homeodoamin simulations}
\label{appendix:homeodomain_analysis}
\begin{figure}
    \centering
    \includegraphics[width=\textwidth]{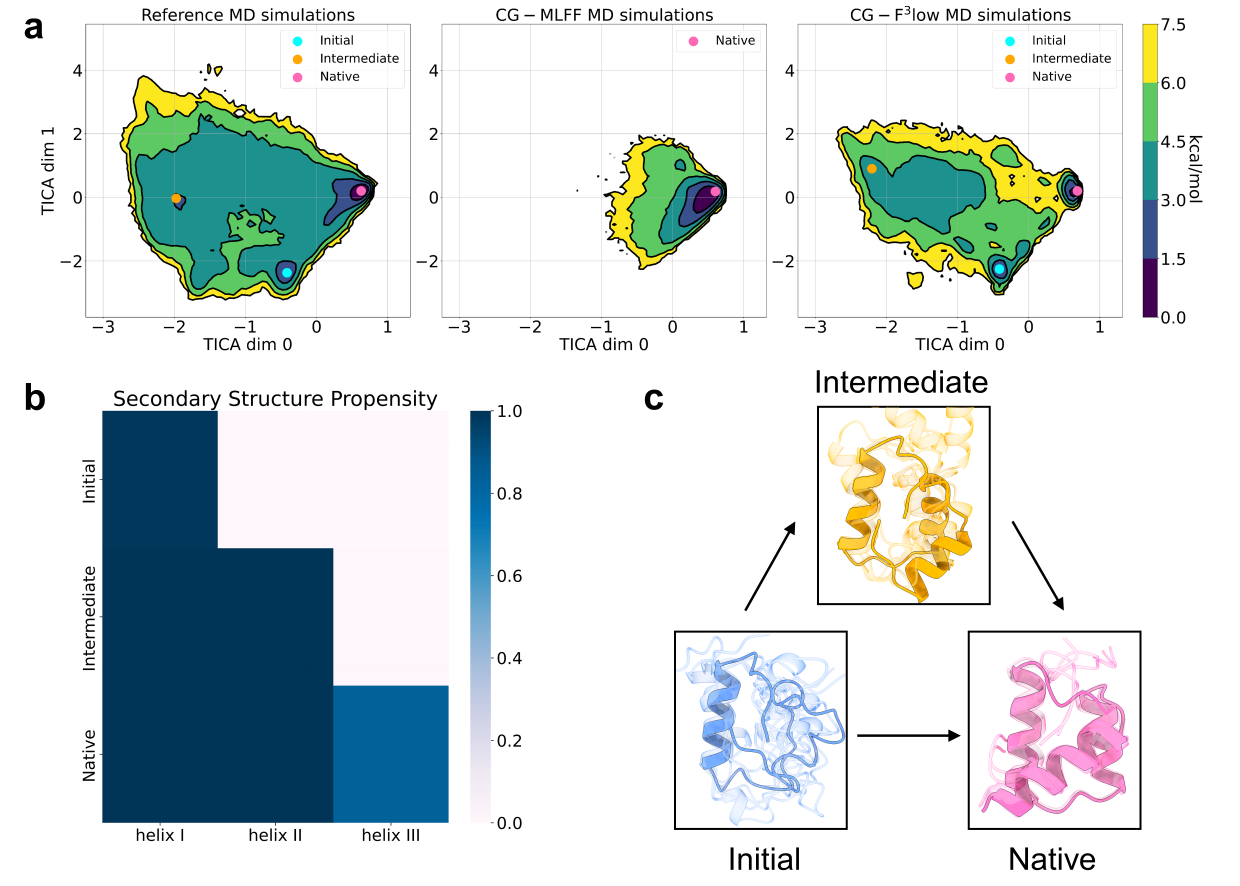}
    \caption{Simulation analysis of Homeodomain. \textbf{a.} The free energy surface of Homeodomain mapped onto the first two principal components derived from tICA for the reference all-atom MD simulations (left), the CG-MLFF simulations (center) and the CG-\model~simulations (right). Distinct macrostates on the landscape are highlighted by circles in different colors: cyan for the initial state, orange for the intermediate state, and pink for the native state. \textbf{b.} The propensity of the three secondary structural elements of the Homeodomain across the macrostates, was measured by the percentage of conformational ensembles with an RMSD threshold of 2$\angstrom$ for each helix. \textbf{c.} The representative conformations from the macrostates identified in the CG-\model~simulations correspond to the free energy minima indicated by the same color coding. Transparent structures reveal additional diverse conformations from the same state. Arrows represent the main pathways transitioning from the initial unstructured state to the native folded state.}
    \label{fig:homeodomain_analysis}
\end{figure}

To demonstrate the capabilities of the \model~in exploring conformational space, we took the Homeodomain as a case study. The conformational landscape explored by the CG-\model~exhibited a similar feature as the reference all-atom MD simulations, successfully identifying three metastable states. On the contrary, the CG-MLFF was confined to a single metastable state (Figure~\ref{fig:homeodomain_analysis}(a)). 
Furthermore, the Homeodomain (54 AA in total) is structured into a three-helix bundle, specifically helix I (8-20 AA), helix II (26-36 AA), and helix III (40-53 AA). 
Notably, helices I and III together form a Helix-Turn-Helix (HTH) motif, a signature motif of DNA-binding proteins and known for its fast-folding properties. 
Experiment results showed that the HTH motif is more prone to instability than helix I, and the unfolding pathway preferentially disrupts the HTH motif before helix I (\cite{mayor2000protein}). 
Our simulations corroborate these observations, further offering insights into its folding pathway (Figure~\ref{fig:homeodomain_analysis}(b) and (c)). 
From the representative conformational ensembles, it is inferred that helix I appears in all states, even in the unstructured initial microstate. Subsequently, it could transition to an intermediate state, wherein the secondary structures of helix I and helix II are nearly formed, and the tertiary interactions are partially established. Such findings support the hypothesis that the Homeodomain employs a diffusion-collision mechanism for folding. 
In the mechanism, the secondary structural elements first form independently and later assemble to the tertiary structure, aligning with prior studies of Homeodomain (\cite{mayor2000protein, demarco2004diffusing}).

\newpage
\section{Experiment details}
\label{appendix:details}

\subsection{Trajectory Dataset}
We adopt the large-scale all-atom molecular dynamics dataset in \cite{majewski2023machine}, which contains twelve fast-folding proteins studied by \cite{lindorff2011fast}.
These proteins exhibit diverse secondary structural elements, including $\alpha$-helices and $\beta$-strands, with lengths spanning from 10 to 80 amino acids.
The evaluation of the proposed methods is conducted on three specific protein trajectories selected from the dataset, which are Chignolin (10AA), Trpcage (20AA), and Homeodomain (54AA).
Other settings are shown in Table~\ref{table:md_dataset}.

\begin{table}[htbp]
\centering
\caption{All-atom MD simulation trajectory dataset from \cite{majewski2023machine}.}  
\resizebox{0.85\textwidth}{!}{
\begin{tabular}{lccc}
\toprule
    Protein & Sequence length (\#AA) & Aggregated time ($\mu s$) & Min. $C_{\alpha}$ RMSD ($\angstrom$) \\
    \midrule
    Chignolin & 10  & 186 & 0.15 \\
    Trpcage & 20 & 195 & 0.45 \\
    Homeodomain & 54 & 198 & 0.56 \\
\bottomrule
\end{tabular}
}
\label{table:md_dataset}
\end{table}

\subsection{Training details}
We employ the FramePred model architecture from \cite{yim2023se}, which stacks multiple Invariant Point Attention layers introduced by AlphaFold2 \cite{jumper2021highly}, which encodes the structural features and decodes the rigid postures in an equivariant way.
Additionally, we augment the model with sequence embedding as single features for input.
All other model hyperparameters follow \cite{yim2023fast}.
For guided flow, we set initial guess weight $\gamma$ to $0.5$, guidance weight $\omega$ to 1 and unconditional probability $p_{\operatorname{uncond}}$ to $0$, in order to conduct a fully guided frame-to-frame simulation.
We would explore the effect of $\gamma$ and $p_{\operatorname{uncond}}$ in the future.
The training configuration includes a maximum of $100$ training epoch, a learning rate of $0.0001$.
For Chignolin and Trpcage, a batch size of $512$ is employed, while for Homeodomain, a batch size of $128$ is utilized.
All training experiments are performed individually on a single NVIDIA GeForce RTX 4090 GPU.

\subsection{Simulation details}

After training, we executed 8 parallel coarse-grained simulations (sampling procedures) with 150,000 samples per protein. 
These simulations were initiated from conformational states that were randomly selected from the reference free energy surface within the test set. 
To remove the effect of the starting conformations, we removed the first 10\% of sample points per trajectory when plotting the free energy surfaces.
All simulation experiments are performed individually on a single NVIDIA
GeForce RTX 4090 GPU.
The simulation time per step is reported in Table \ref{table:sim_time}. 

\begin{table}[htbp]
    \centering
    \begin{tabular}{cccc}
    \toprule
         &  Chignolin &  Trpcage & Homeodomain \\
         \midrule
        Simulation time per step (s) & 0.32 $\pm$ 0.01  & 0.34 $\pm$ 0.01 & 0.38 $\pm$ 0.01 \\
    \bottomrule
    \end{tabular}
    \caption{Simulation time per step for each protein.}
    \label{table:sim_time}
\end{table}

\end{document}

%% file: math_commands.tex
\usepackage{amsmath,amsfonts,bm}

\def\eqref#1{equation~\ref{#1}}

\def\1{\bm{1}}

\DeclareMathAlphabet{\mathsfit}{\encodingdefault}{\sfdefault}{m}{sl}
\SetMathAlphabet{\mathsfit}{bold}{\encodingdefault}{\sfdefault}{bx}{n}

